\documentclass[review]{elsarticle}

\usepackage{amssymb} 

\usepackage{lineno,hyperref}
\modulolinenumbers[5]

\journal{Optik}

\newcommand{\sech}{\mbox{sech}}

\usepackage{amsmath}
\usepackage{mathtools}
\usepackage{float}

\begin{document}

\begin{frontmatter}

\title{Management of solitons in medium with competing cubic and quadratic nonlinearities}

\author{F. Kh. Abdullaev$^{1,2}$}
\author{J. S. Yuldashev$^{1,2}$}
\author{M. \"{O}gren$^{3,4}$}
\address{$^1$ Physical-Technical Institute, Uzbek Academy of Sciences, 100084 Tashkent, Uzbekistan}
\address{$^2$ Theoretical Physics Department, National University of Uzbekistan, 100174 Tashkent, Uzbekistan}
\address{$^3$School of Science and Technology, \"{O}rebro University, 70182 \"{O}rebro, Sweden}
\address{$^4$Hellenic Mediterranean University, P.O. Box 1939, GR-71004, Heraklion, Greece}




\begin{abstract}
Management of solitons in media with competing quadratic and cubic nonlinearities is investigated. 
Two schemes, using rapid modulations of a mismatch parameter, and of the Kerr nonlinearity parameter are studied. 
For both cases, the averaged in time wave equations are derived. 
In the case of mismatch management, the region of the parameters where stabilization is possible is found. 
In the case of Kerr nonlinearity management, it is shown that the effective $\chi^{(2)}$ nonlinearity depends on the intensity imbalance between fundamental (FH) and second (SH) harmonics. Predictions obtained from the averaged equations are confirmed by numerical simulations of the full PDE's.
\end{abstract}

\begin{keyword}
Solitons, mismatch management, Kerr nonlinearity management, competing nonlinearities.
\end{keyword}

\end{frontmatter}




\section{Introduction}\label{sec_Int}
Solitons in media with quadratic ($\chi^{(2)}$) and cubic ($\chi^{(3)}$) nonlinearaties have been studied intensively experimentally as well as theoretically.
Special interest is attracted to optical dispersive media with both $\chi^{(2)}$ and $\chi^{(3)}$ nonlinearities present,
i.e., to the case of so-called competing nonlinearities.
Such systems also appears from the mean-field description of matter waves in atomic-molecular condensates (AMBEC).
In the latter case the role of cubic nonlinearities is played by the two-body interactions for atomic and molecular fields, respectively.
The role of the quadratic nonlinearities is played by conversion of pairs of atoms into molecules and vice versa.
The analysis of the existence of optical solitons and their stability in such systems has been performed  in
\cite{Buryak1}, and for the case of matter waves in~\cite{Sacha}.
In particular it have been shown in~\cite{BuryakKivshar} that stable bright solitons can exists in a media with competing nonlinearities, and that the strength of the $\chi^{(2)}$ non-linearity plays an important role for the instability  of the solitons.

Hence, it is of importance to develope methods for stabilization and control of solitons in these systems. 
The method of dynamical stabilization of solitons by using rapid variations in time of parameters in the media have been suggested in~\cite{SaitoPRL2003}.
Reviews of these results are contained in the books~\cite{Malomed1, Biswas}, and a more recent review is~\cite{Malomed2}.

One of the possible methods of dynamical stabilization can be mismatch management. 
Mismatch management in a pure $\chi^{(2)}$ system has been considered in \cite{Driben, Matus} where regions of stability and instability has been analyzed for the slowly varying and resonant cases mainly, and in~\cite{CB,Conforti} for quasi-phase-matching (QPM) regime.
Also the influence of the management of the cubic- and quadratic-nonlinearities on soliton stability and continuous wave (CW) dynamics have been studied in~\cite{Matus,AMY}, where the resonant responses of solitons and CW have been analyzed.

Here we will study the possible roles of the rapidly varying periodic mismatch and the Kerr non-linearity for the dynamical stabilization and the control of optical and matter-wave solitons in media with competing cubic- and quadratic-nonlinearities.

The structure of the article is the following:

In section~2 we describe the model used for the propagation of the waves in media with competing cubic and quadratic non-linearities. 
The system obtained by averaging over rapid modulations of the mismatch parameter, and a condition for dynamical stabilization of solitons, are obtained in section~3. 
The averaged equations and solitons for a system with a periodically modulated Kerr non-linearity parameter, and the corresponding full numerical simulations, are considered in section~4.
Finally, we present the conclusions of the investigations in section~5.

\section{The model} \label{sec_intro}

The system, describing the propagation of the fundamental- (FH) and second- (SH) harmonics in a quadratic nonlinear media with a cubic
nonlinearity have in standard optics dimensionless variables the form~\cite{BuryakKivshar}
\begin{eqnarray}\label{model}
i\frac{\partial u}{\partial z} + r\frac{\partial ^2 u}{\partial x^2} - u + u^* w + \chi \left(\frac{1}{2\sigma}\lvert u\rvert^2 + \rho \lvert w\rvert^2\right)u=0,  \nonumber \\
i\sigma \frac{\partial w}{\partial z} + s\frac{\partial^2 w}{\partial x^2} - qw + \frac{1}{2}u^2 + \chi\left(2\sigma\lvert w\rvert^2 + \rho \lvert u\rvert^2\right)w=0.
\end{eqnarray}
Here $u,w$ are the fields of the FH and the SH respectively.
The parameter $\chi \sim \chi^{(3)}/\chi^{(2)^2} $ characterize a balance of contributions that are due to the $\chi^{(2)}$ and $\chi^{(3)}$ nonlinearities.
We consider spatial bright solitons ($r=s=+1$) and select the same parameter values ($\rho=\sigma=2$) as in~\cite{BuryakKivshar}.
Below we consider two cases separately, a longitudinally modulated linear mismatch parameter, $q(z)$, and a modulated cubic nonlinearity, $\chi(z)$. 
We decompose the modulation of the mismatch parameter $q(z)$ (cubic nonlinearity $\chi(z)$), into a mean-value part $q_0$ ($\chi_0$), and a fastly varying part $q_1(z)$ ($\chi_1(z)$) with a large amplitude, as follows:
\begin{equation}\label{modulated_parameters}
q(z)= q_0 + q_1(z)=q_0 + q_1\cos(\omega z), \,\,\,
\Big(\chi(z) = \chi_0 + \chi_1(z) = \chi_0 + \chi_1\cos(\omega z)\Big).
\end{equation}

\section{Longitudinally modulated mismatch parameter}
First we obtain the averaged equations in the case of $q \rightarrow q(z)$ and $\chi = \chi_0$. We consider the two cases of a small parameter mismatch and a large parameter mismatch. 
In the former case (small $q_0$) we use the following transformation for the second-harmonic field for deriving so-called averaged equations
\begin{equation}\label{transformations_small_mismatch}
w = ve^{-i\frac{\varGamma(z)}{\sigma}},
\end{equation}
where $\varGamma_z(z)=q_1(z)$, i.e., from Eq.~(\ref{modulated_parameters}) we have $\varGamma(z)=\frac{q_1}{\omega}\sin(\omega z)$.
This transformation permits us to eliminate the strong rapid varying terms~\cite{SaitoPRL2003}.
By using mismatch parameters in the form of~(\ref{modulated_parameters}) and substituting the new second-harmonic field $v$ into Eqs.~(\ref{model}) we get similar equations to Eqs.~(\ref{model}), but with the quadratic nonlinear terms with exponential factors
\begin{eqnarray}\label{eq4}
i\frac{\partial u}{\partial z} + \frac{\partial^2 u}{\partial x^2} - u + u^*ve^{-\frac{i}{\sigma}\varGamma(z)} + \chi_0\left(\frac{1}{4}\lvert u\rvert^2 + 2 \lvert v\rvert^2\right)u&=&0,  \nonumber \\
i \frac{\partial v}{\partial z} + \frac{1}{2}\frac{\partial^2 v}{\partial x^2} - \frac{q_0}{2} v + \frac{1}{4}u^2e^{\frac{i}{\sigma}\varGamma(z)} + \chi_0\left(2\lvert v\rvert^2 + \lvert u\rvert^2\right)v&=&0.
\end{eqnarray}

We then expand the exponential factors in Eqs.~(\ref{eq4}) into Fourier series and restrict the expansion coefficients to the average values (\textit{zero-order term}) in obtaining an averaged equation, i.e.
\begin{equation}\label{Fourier_expansion}
e^{\pm i\frac{q_1}{\sigma \omega}\sin(z^\prime)}=\sum_{n=-\infty}^{\infty}c_n^\pm e^{inz^\prime} = \sum_{n=-\infty}^{\infty}J_n\left(\pm \frac{q_1}{\sigma \omega}\right)e^{inz^\prime},
\end{equation}
where $z^\prime=\omega z$.
The averaged equations can now be written in the final form:
\begin{eqnarray}\label{averaged_equations_1}
i\frac{\partial u}{\partial z} + \frac{\partial^2 u}{\partial x^2} - u + u^*vJ_0\left(\frac{q_1}{\sigma \omega}\right) + \chi_0\left(\frac{1}{4}\lvert u\rvert^2 + 2 \lvert v\rvert^2\right)u&=&0,  \nonumber \\
i \frac{\partial v}{\partial z} + \frac{1}{2} \frac{\partial^2 v}{\partial x^2} - \frac{q_0}{2} v + \frac{1}{4}u^2J_0\left(\frac{q_1}{\sigma \omega}\right) + \chi_0\left(2\lvert v\rvert^2 + \lvert u\rvert^2\right)v&=&0,
\end{eqnarray}
where $J_0(\cdot)$ is the zero-order Bessel function.

In the latter case (large $q_0$) we use instead a similar transformation, including the large mismatch parameter in the exponential, as was done for the quasi-phase-matching (QPM) scheme in~\cite{Fejer}
\begin{equation}\label{transformations_large_mismatch}
w = ve^{-i\frac{\varGamma(z)}{\sigma}-i\frac{q_0}{\sigma} z}.
\end{equation}
By using the transformation (\ref{transformations_large_mismatch}), we get the following equations

\begin{eqnarray}\label{eq6}
i\frac{\partial u}{\partial z} + \frac{\partial^2 u}{\partial x^2} - u + u^*ve^{-\frac{i}{\sigma}\varGamma(z)}e^{-i\frac{q_0}{\sigma}z} + \chi_0\left(\frac{1}{4}\lvert u\rvert^2 + 2 \lvert v\rvert^2\right)u&=&0,  \nonumber \\
i \frac{\partial v}{\partial z} + \frac{1}{2}\frac{\partial^2 v}{\partial x^2} + \frac{1}{4}u^2e^{\frac{i}{\sigma}\varGamma(z)}e^{i\frac{q_0}{\sigma}z} + \chi_0\left(2\lvert v\rvert^2 + \lvert u\rvert^2\right)v&=&0.
\end{eqnarray}
 We now do the same mathematical procedure as in deriving Eq.~(\ref{averaged_equations_1}), but we also take into account first order terms ($c_1$ and $c_{-1}$) in the Fourier expansion of Eq.~(\ref{Fourier_expansion}), that is
\begin{eqnarray}\label{eq9}
i\frac{\partial u}{\partial z} + \frac{\partial^2 u}{\partial x^2} - u + u^*v\left[J_0\left(\frac{q_1}{\sigma \omega}\right)-J_1\left(\frac{q_1}{\sigma \omega}\right)e^{i\omega z}+J_1\left(\frac{q_1}{\sigma \omega}\right)e^{-i\omega z}\right]e^{-i\frac{q_0}{\sigma}z} \nonumber \\
+ \chi_0\left(\frac{1}{4}\lvert u\rvert^2 + 2 \lvert v\rvert^2\right)u=0,  \nonumber \\
i \frac{\partial v}{\partial z} + \frac{1}{2}\frac{\partial^2 v}{\partial x^2} + \frac{1}{4}u^2\left[J_0\left(\frac{q_1}{\sigma \omega}\right)+J_1\left(\frac{q_1}{\sigma \omega}\right)e^{i\omega z}-J_1\left(\frac{q_1}{\sigma \omega}\right)e^{-i\omega z}\right]e^{i\frac{q_0}{\sigma}z} \nonumber \\
+ \chi_0\left(2\lvert v\rvert^2 + \lvert u\rvert^2\right)v=0,
\end{eqnarray}
where $J_1(\cdot)=-J_{-1}(\cdot)$ is the first-order Bessel function. 
To obtain the averaged equation we omit the fast oscillating terms (with large exponentials) in Eq.~(\ref{eq9})

\begin{eqnarray}\label{eq10}
i\frac{\partial u}{\partial z} + \frac{\partial^2 u}{\partial x^2} - u - u^*vJ_1\left(\frac{q_1}{\sigma \omega}\right)e^{-i(\frac{q_0}{\sigma}-\omega)z} + \chi_0 \left( \frac{1}{4}\lvert u\rvert^2 + 2 \lvert v\rvert^2 \right)u&=&0,  \nonumber \\
i \frac{\partial v}{\partial z} + \frac{1}{2}\frac{\partial^2 v}{\partial x^2} - \frac{1}{4}u^2J_1\left(\frac{q_1}{\sigma \omega}\right)e^{i(\frac{q_0}{\sigma}-\omega)z} + \chi_0\left(2\lvert v\rvert^2 + \lvert u\rvert^2\right)v&=&0.
\end{eqnarray}

By applying the following transformation for the second-harmonic: $v \rightarrow \bar{v}e^{i\delta q z}$, we can again write averaged equations similar to Eq.~(\ref{model}), but with re-normalized terms
\begin{eqnarray}\label{averaged_equations_2}
i\frac{\partial u}{\partial z} + \frac{\partial^2 u}{\partial x^2} - u - u^*vJ_1\left(\frac{q_1}{\sigma \omega}\right) + \chi_0\left(\frac{1}{4}\lvert u\rvert^2 + 2 \lvert v\rvert^2\right)u&=&0,  \nonumber \\
i \frac{\partial v}{\partial z} + \frac{1}{2}\frac{\partial^2 v}{\partial x^2} - \delta q v - \frac{1}{4}u^2 J_1\left(\frac{q_1}{\sigma \omega}\right) + \chi_0\left(2\lvert v\rvert^2 + \lvert u\rvert^2\right)v&=&0,
\end{eqnarray}
where $\delta q = q_0 / \sigma - \omega$, and we have dropped the bar-sign for convenience.

Stability of soliton solutions of Eqs.~(\ref{model}) in the case of small mismatch parameters ($q_0=2$ and $q_1=0$) was investigated numerically in~\cite{BuryakKivshar}, and the three most physically important soliton families of low energy were presented.
Here the energy is defined by~\cite{BuryakKivshar}
\begin{equation} \label{eq:def:energy}
P = |\chi_0|\int \left( |u|^2 +4|w|^2 \right)dx.
\end{equation}
According to that investigation, solitons with energy higher than $8\sigma$ ($W$ -\textit{ type solitons}) are unstable for any value of the parameter $\chi_0$,  while solitons $v(x)=\sqrt{q_0/(\sigma \chi_0)}\sech(\sqrt{q_0}x)$ with energy $4\sqrt{2}q_0/\sigma$ ($V$ - \textit{type solitons}) are stable for $\chi_0 \geq 8.76$, but unstable for $0< \chi_0 \lessapprox 8.76$.
It is apparent from the averaged Eqs.~(\ref{averaged_equations_1}) and ~(\ref{averaged_equations_2}) that the contribution of the quadratic nonlinearities can be controlled by changing the parameters $q_1$ or $\omega$.
Therefore, stabilization of $W$ and $V$ type solitons may be possible as we can reduce the effect of the quadratic nonlinearity by changing modulation parameters in the unstable range of $\chi_0$.

\begin{figure}[H]
\begin{center}
\includegraphics[scale=0.18]{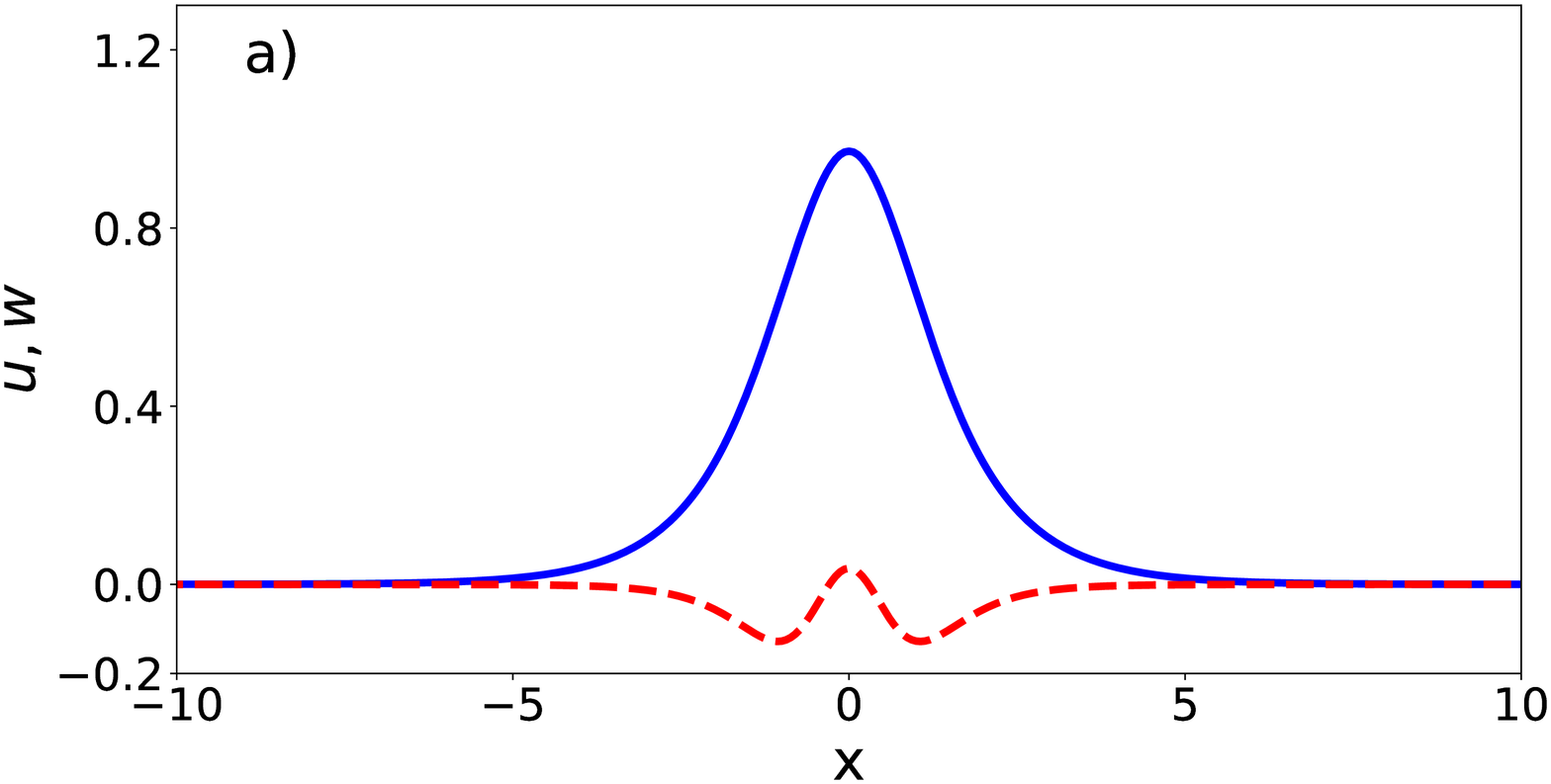} 
\includegraphics[scale=0.18]{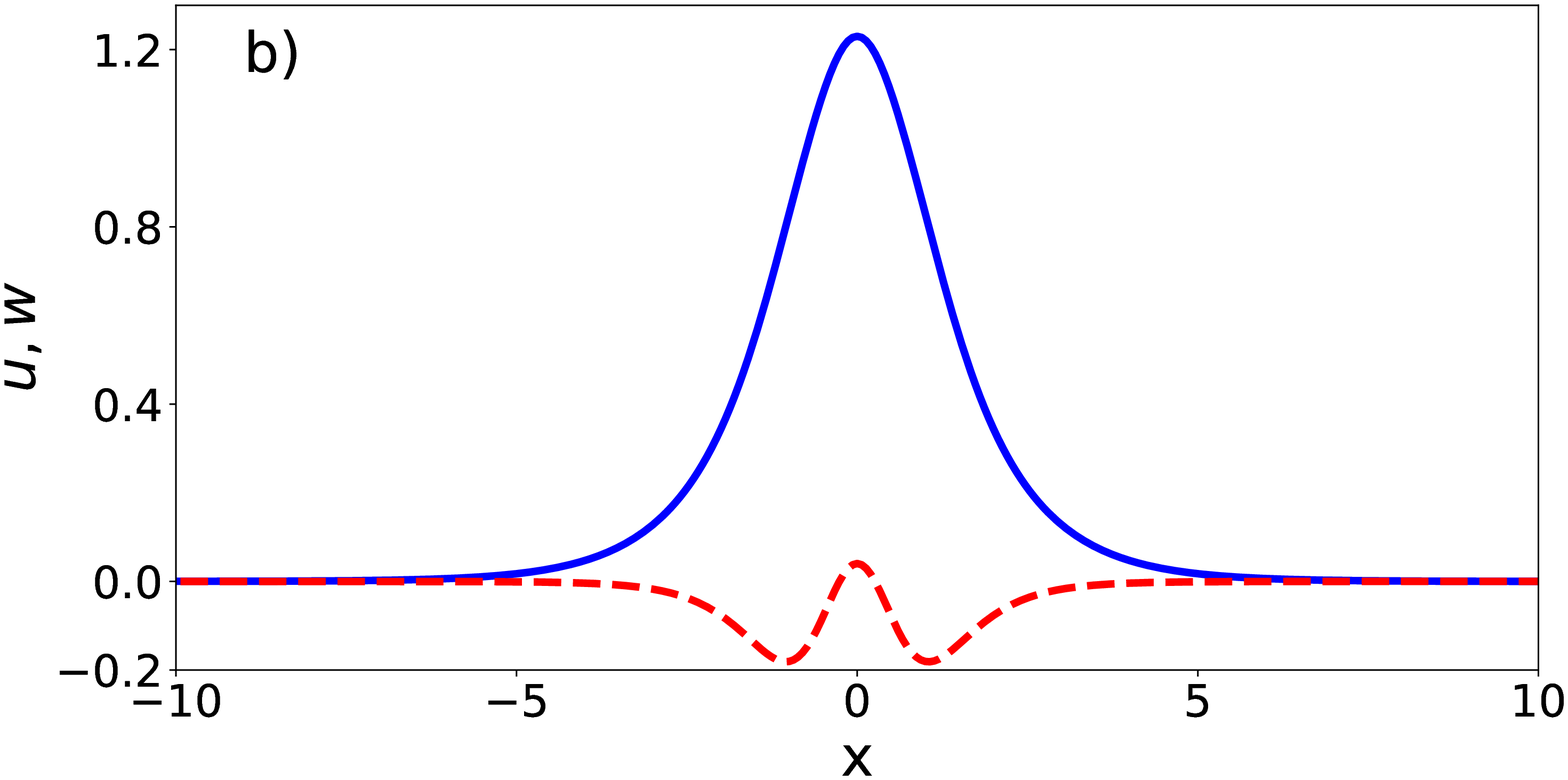}
\caption{(Color online)
$W$ - type solitons, stationary solutions of Eqs.~(\ref{model}).
a) Soliton profile with total energy $P = 16.8182$ and Kerr parameter $\chi_0=8$;
b) Soliton profile with total energy $P = 17.0947$ and Kerr parameter $\chi_0=5$.
Solid (blue) curves corresponds to FH and dashed (red) curves corresponds to SH. 
The other parameters, $q_0=2$ and $q_1 = 0$, are the same for both cases.}
\label{fig:1}
\end{center}
\end{figure}

\subsection{Numerical analysis}
\begin{figure}[H]
\begin{center}
\includegraphics[scale=0.25]{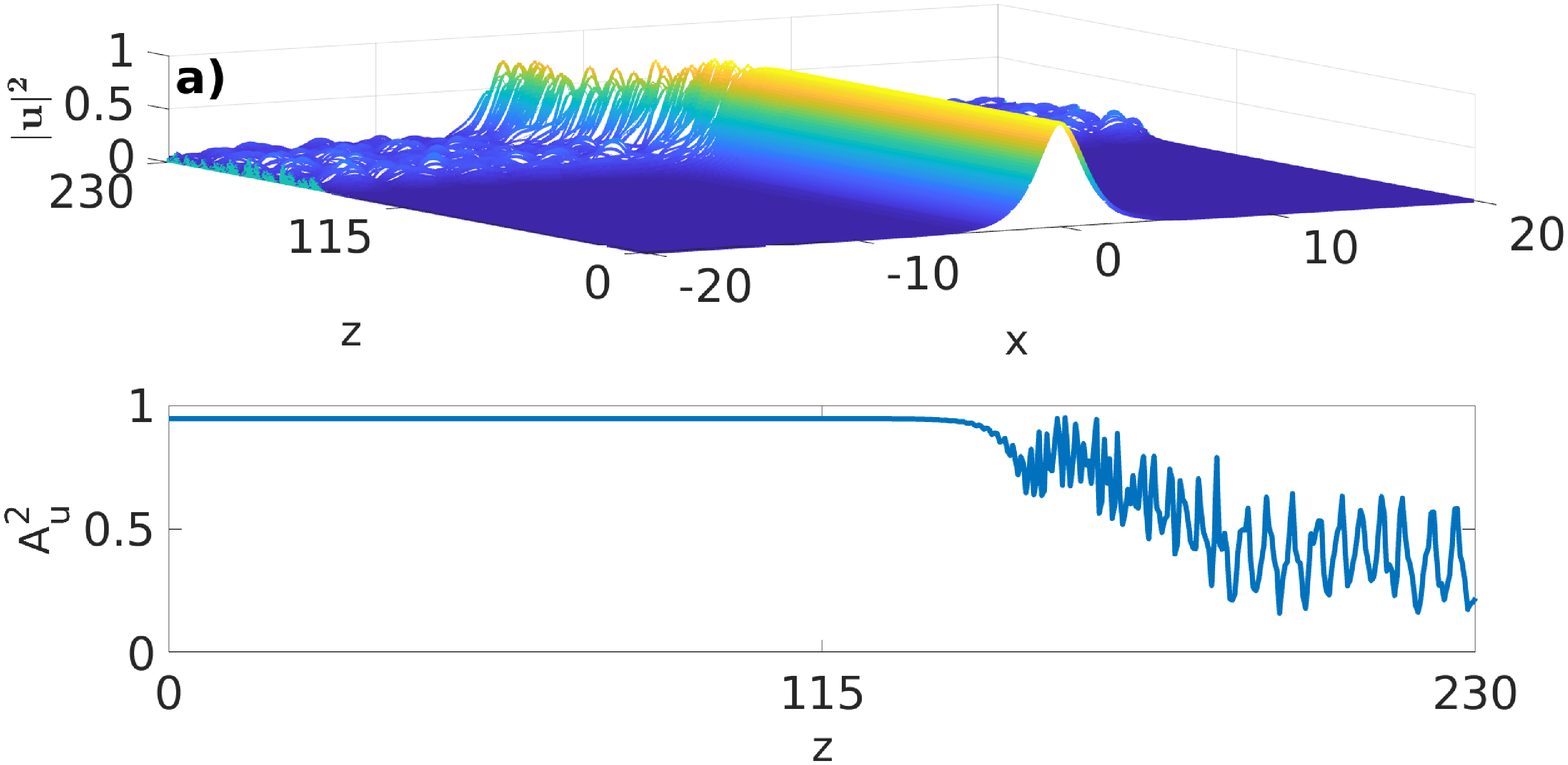}
\includegraphics[scale=0.25]{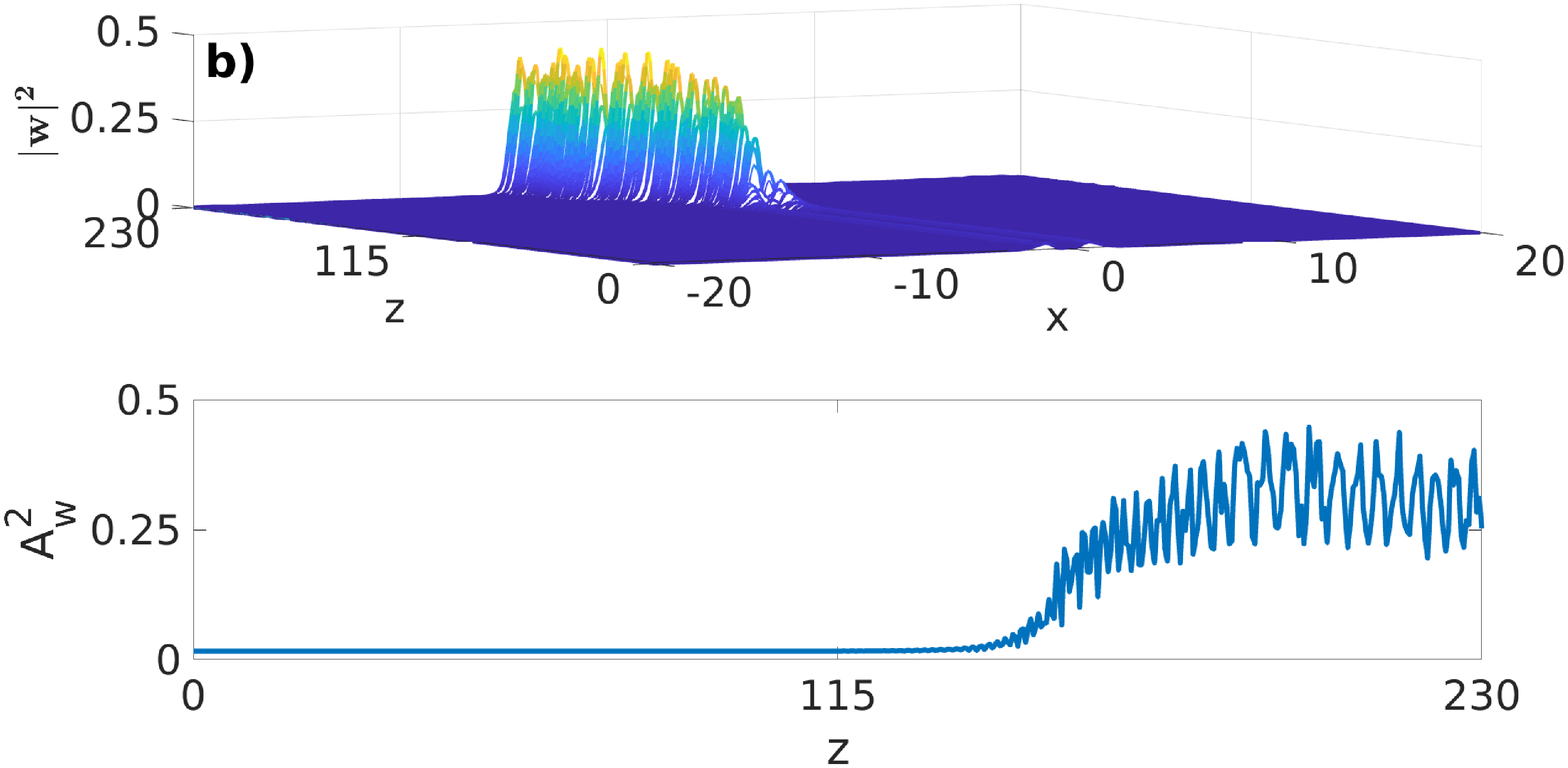}
\includegraphics[scale=0.25]{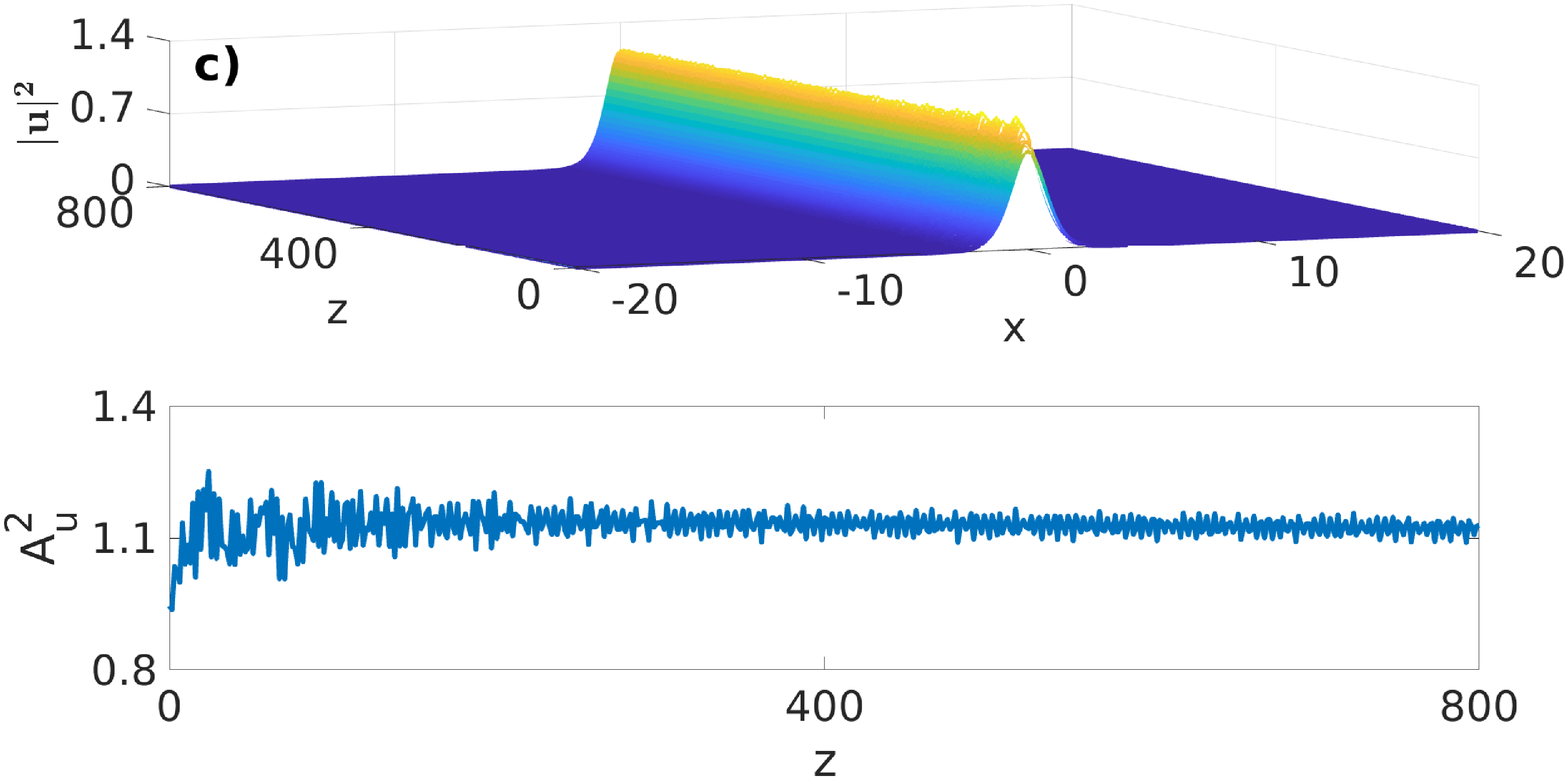}
\includegraphics[scale=0.25]{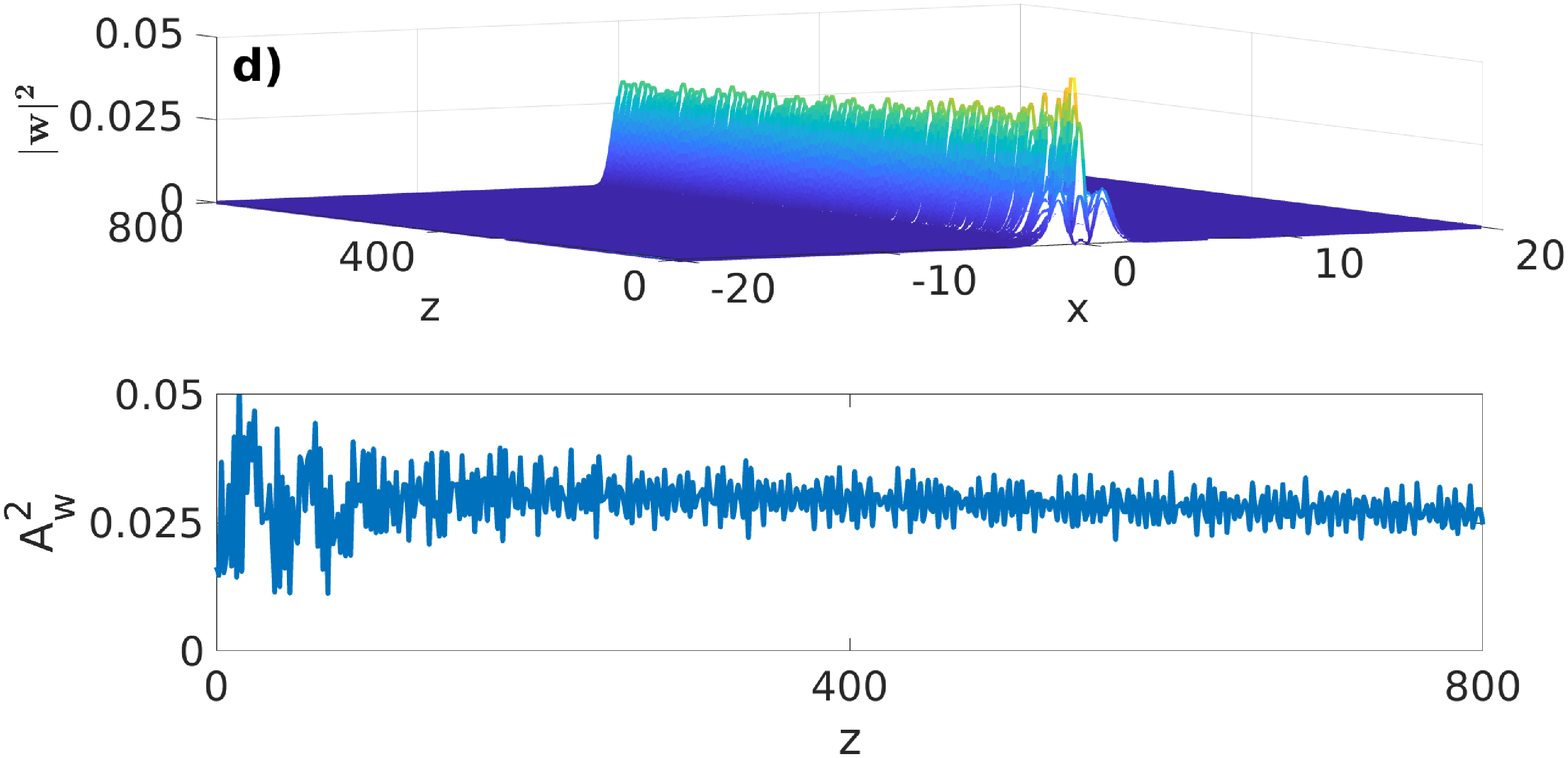}
\caption{(Color online) Evolution of intensities ($|u|^2$, $|w|^2$) and the squared central amplitudes ($A^2_u$, $A^2_w$) for the stationary solution [Fig.~\ref{fig:1} a)] in the unstable region for $\chi_0=8$. Frames a) and b) demonstrate the unstable dynamics of the fundamental- and second-harmonics waves and their amplitudes, respectively, when the mismatch parameter $q$ is not modulated ($q_1=0$). Frames c) and d) correspond to stabilization of these solitons when the mismatch parameter $q(z)$ is modulated ($q_1=96.2$, $\omega=20$).}
\label{fig:2}
\end{center}
\end{figure}

\begin{figure}[H]
\begin{center}
\includegraphics[scale=0.25]{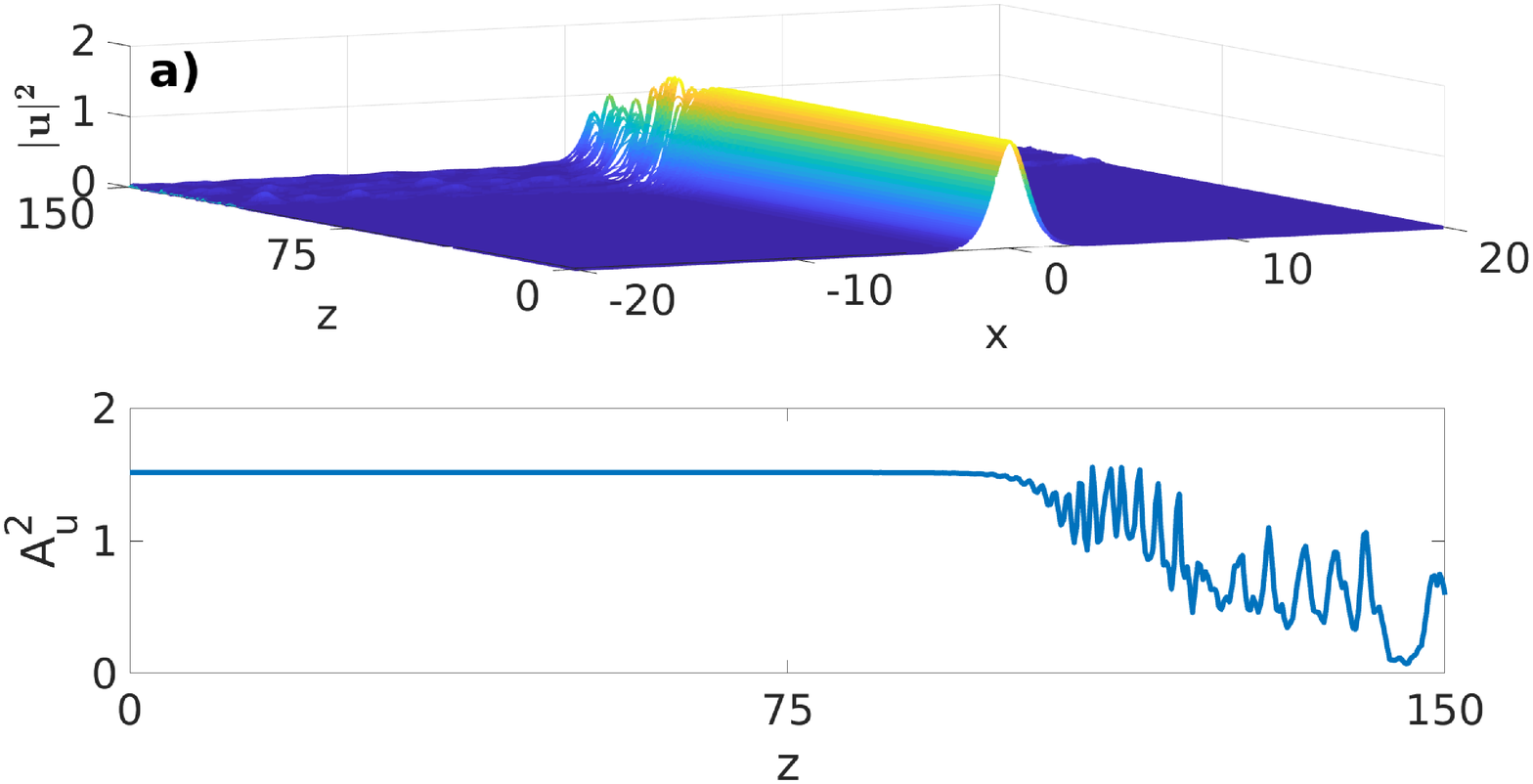}
\includegraphics[scale=0.25]{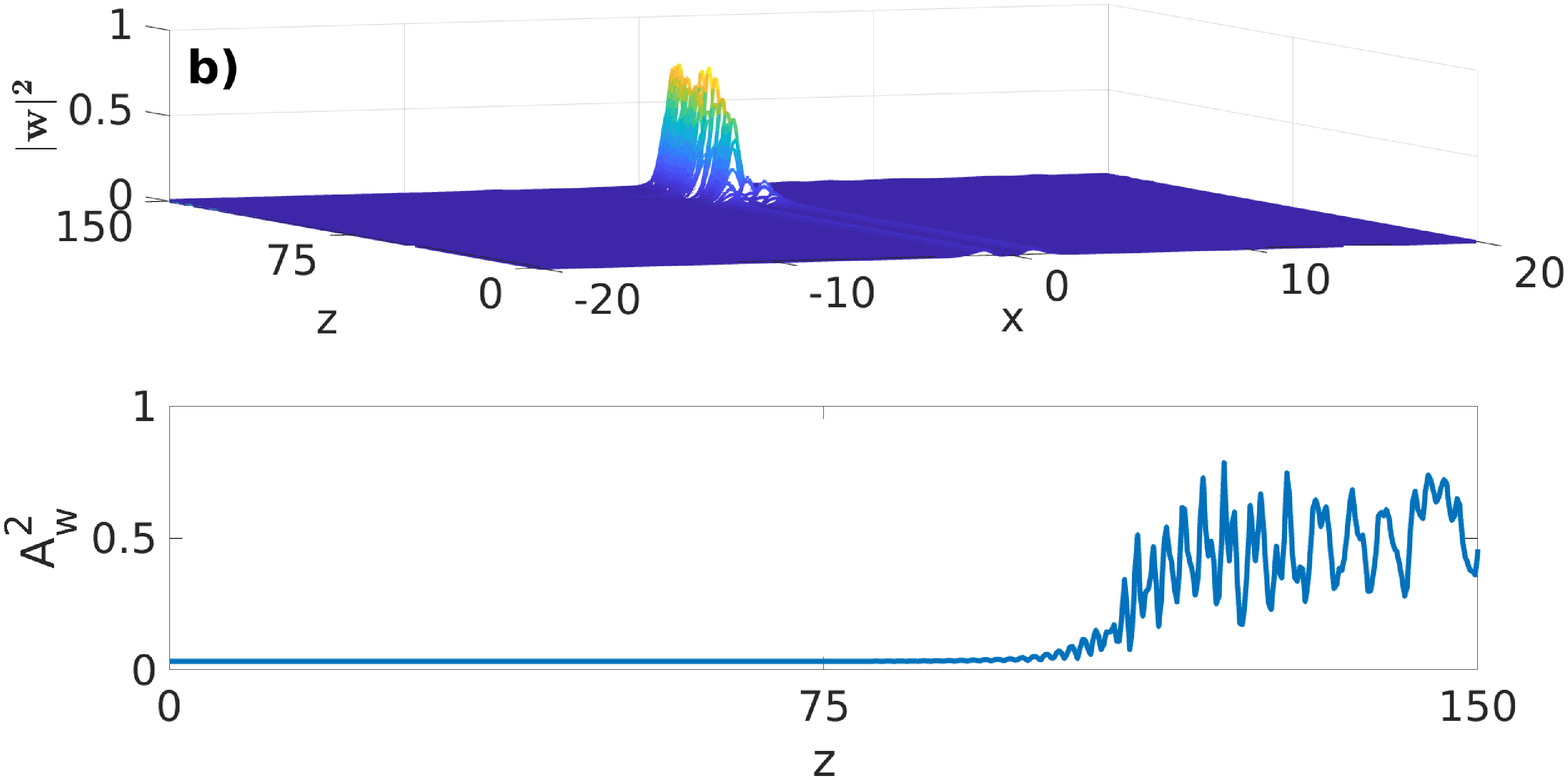}
\includegraphics[scale=0.25]{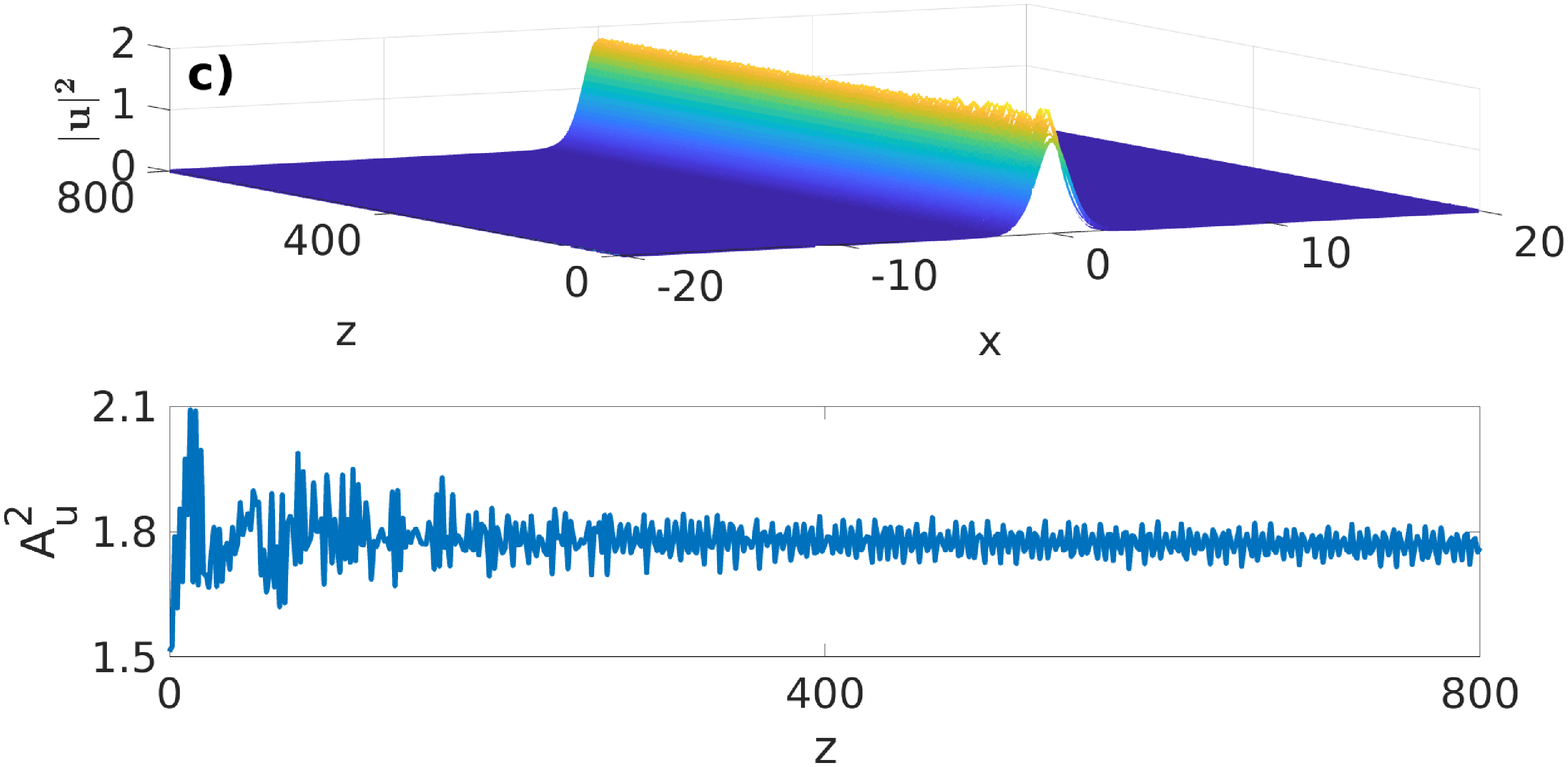}
\includegraphics[scale=0.25]{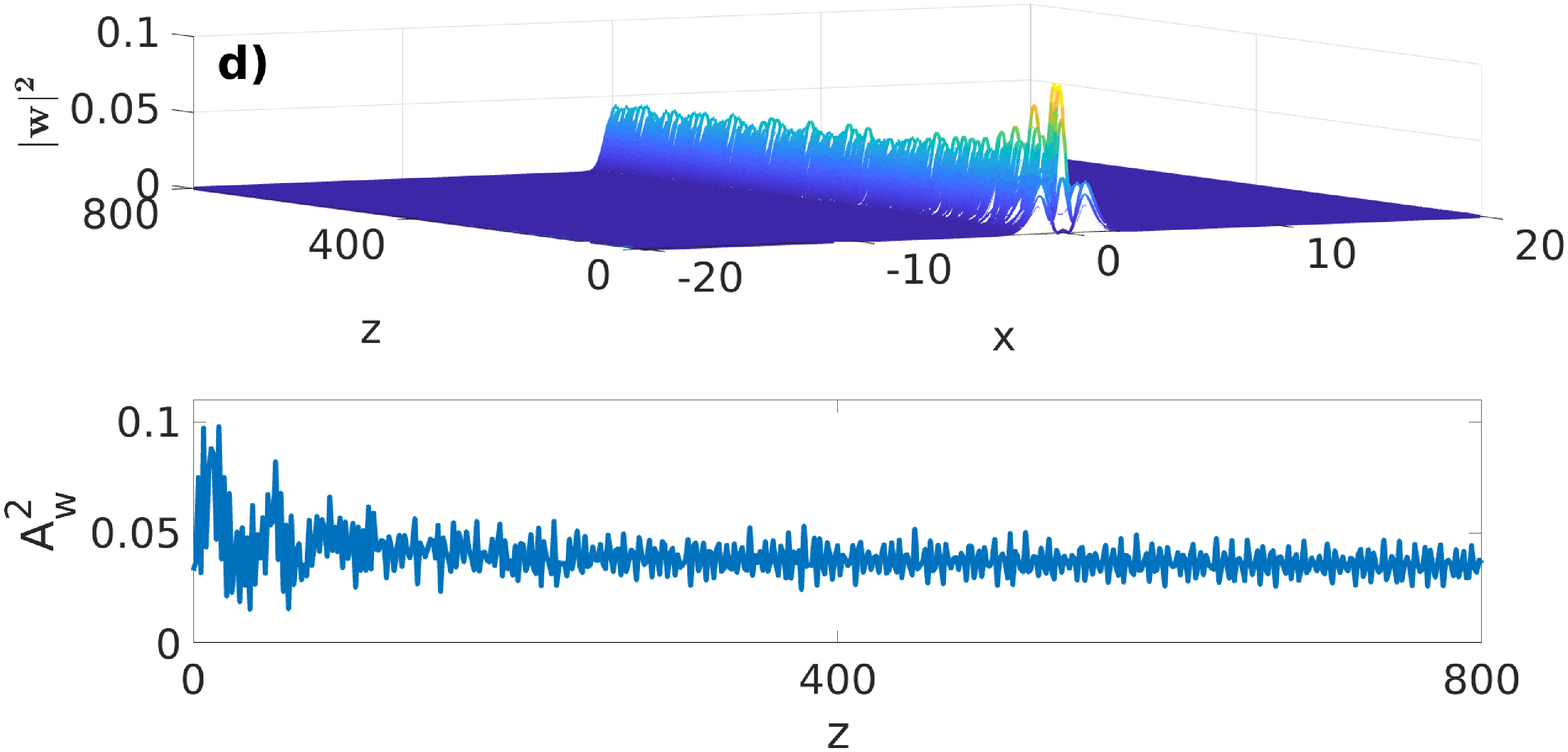}
\caption{(Color online) Evolution of intensities ($|u|^2$, $|w|^2$) and the squared central amplitudes ($A^2_u$, $A^2_w$) for the stationary solution [Fig.~\ref{fig:1} b)] in the unstable region for $\chi_0=5$. Frames a) and b) demonstrate the unstable dynamics of the fundamental- and second-harmonics waves and their amplitudes, respectively, when the mismatch parameter $q$ is not modulated ($q_1=0$). Frames c) and d) correspond to stabilization of these solitons when the mismatch parameter $q(z)$ is modulated ($q_1=96.2$, $\omega=20$).}
\label{fig:3}
\end{center}
\end{figure}

\begin{figure}[H]
\begin{center}
\includegraphics[scale=0.23]{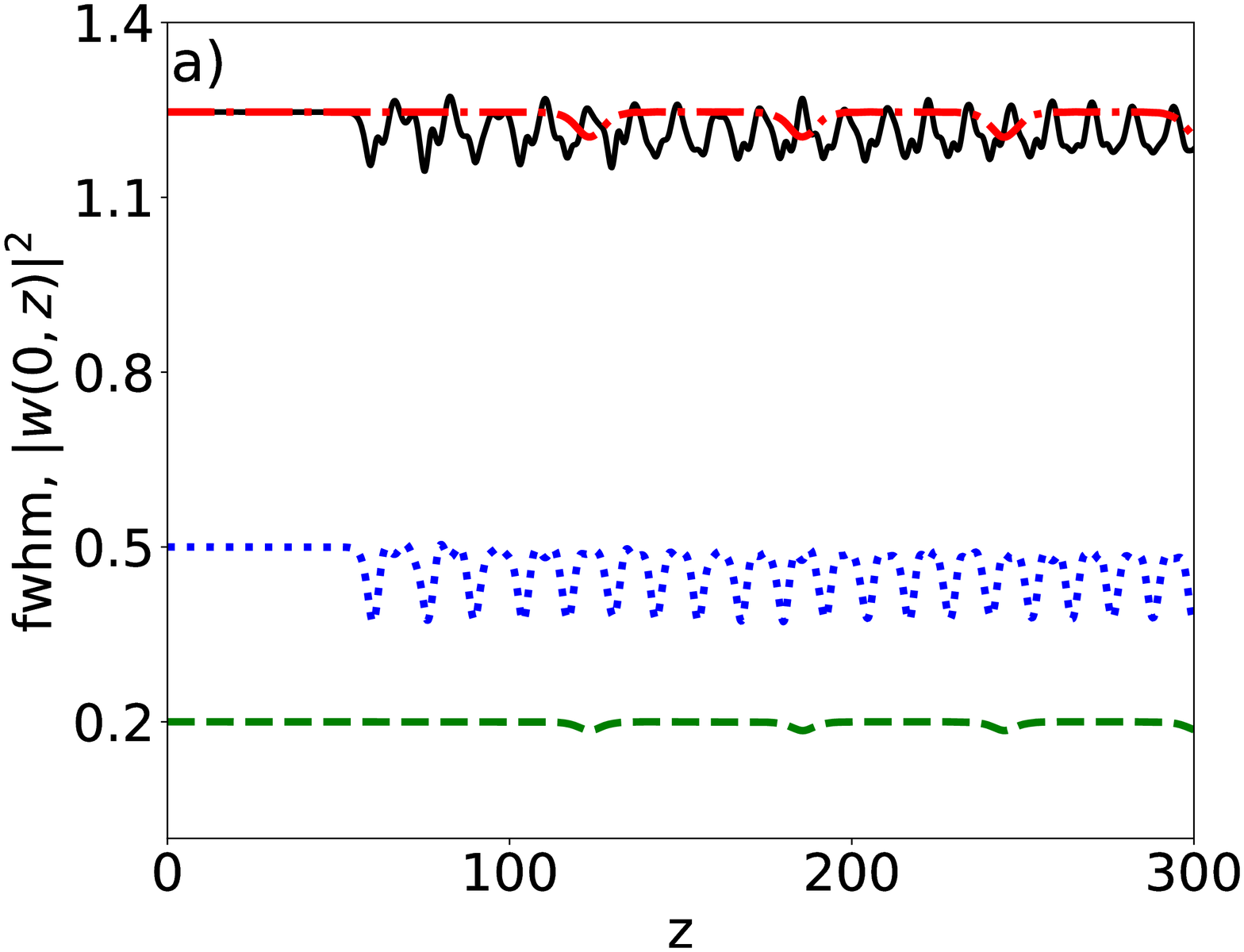}
\includegraphics[scale=0.23]{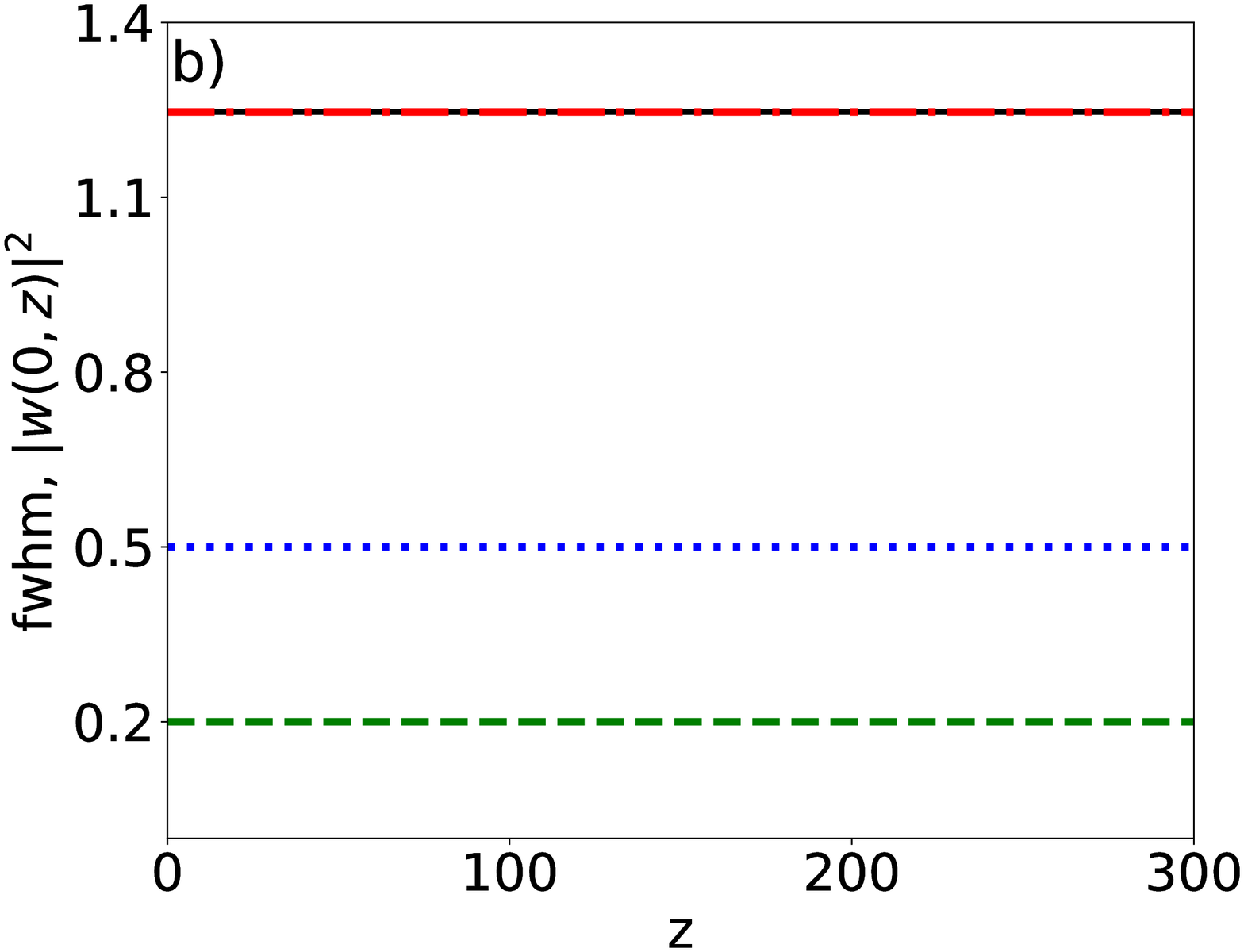}

\caption{(Color online) Evolution of width and the squared central amplitude for stationary solutions with and without modulation of the mismatch parameter $q$. Frame a) demonstrate the unstable dynamics of the $V$ - type soliton in the case of $\chi_0=2$, solid (black) and dotted (blue) curves correspond to soliton's full width at half maximum ({\it fwhm}) and the squared central amplitude ($|w(0,z)|^2$), respectively, of the second-harmonic; and for $\chi_0=5$ dash-dotted (red) and dashed (green) curves correspond to the soliton's full width at half maximum ({\it fwhm}) and the squared central amplitude ($|w(0,z)|^2$), respectively, when the mismatch parameter is not modulated ($q_1=0$). Frame b) corresponds to stabilization of the solitons for the same stationary solutions when the mismatch parameter $q(z)$ of Eq.~(\ref{modulated_parameters}) is modulated with the frequency $\omega=10$ ($q_1=32$: black and blue curves, and $q_1=25$: red and green curves). In both frames, $q_0=2$.}
\label{fig:4}
\end{center}
\end{figure}

We have carried out numerical calculations to check the stability of $W$ and $V$ type solitons for different parameter values of $\chi_0$ and the energies, for which these solitons are unstable.
It has been noted that $W$ - type solitons are supported mainly by the cubic nonlinearity, but the second-harmonic, which has two bumps in the profile as shown in Fig.~\ref{fig:1}, is also generated as the result of parametric coupling. 
The numerical analysis of the original equations~(\ref{model}), shows that modulation of the mismatch parameter along the direction of propagation leads to stabilization of the unstable $W$ - type (Figs.~\ref{fig:2} and~\ref{fig:3}), and of the $V$- type solitons (Fig.~\ref{fig:4}). 
In Fig.~\ref{fig:2} and Fig.~\ref{fig:3}, the dynamics of the intensity profiles ($|u(x,z)|^2$ and $|w(x,z)|^2$) and the squared central amplitudes ($A_u^2=|u(0,z)|^2$ and $A_w^2=|w(0,z)|^2$) of the stationary states [Fig.~\ref{fig:1}, a) and b)] of Eqs.~(\ref{model}) in the case of $\chi_0=8$ and $\chi_0=5$, with total energy $P = 16.8182$ and $17.0947$, respectively, are shown.

For the $V$ - type solitons with small mismatch parameter, numerical simulations of the full equations~(\ref{model}), shows that stabilization of stationary states in the unstable region, $0< \chi_0 \lessapprox 8.76$ and $P = 4\sqrt{2}$, can be achieved by modulating the mismatch parameter with $q_1 = 32$, and $q_1=25$, in the case of $\chi_0=2$, and $\chi_0=5$, respectively, see Fig.~\ref{fig:4}.

\begin{figure}[H]
\begin{center}
\includegraphics[scale=0.23]{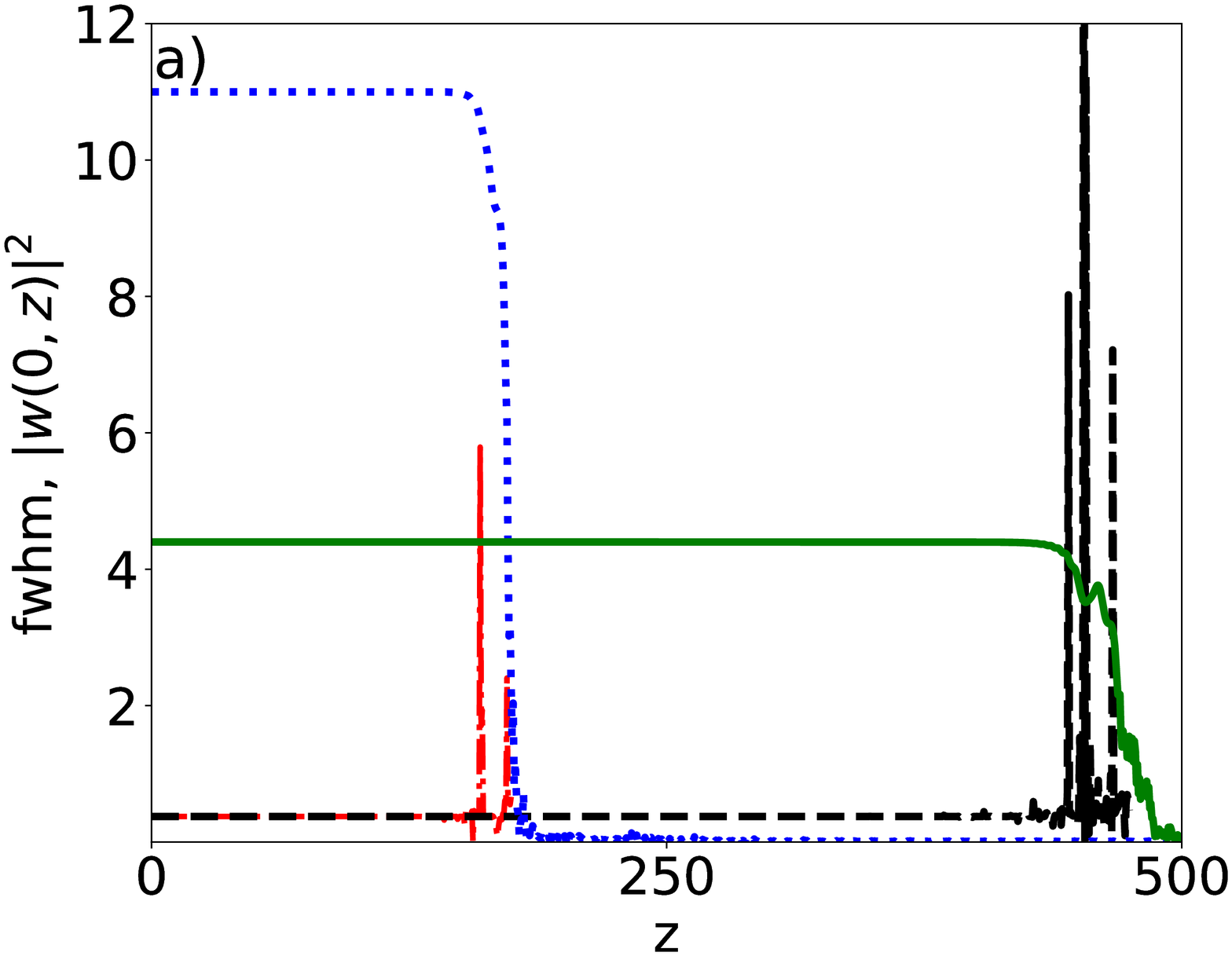}
\includegraphics[scale=0.23]{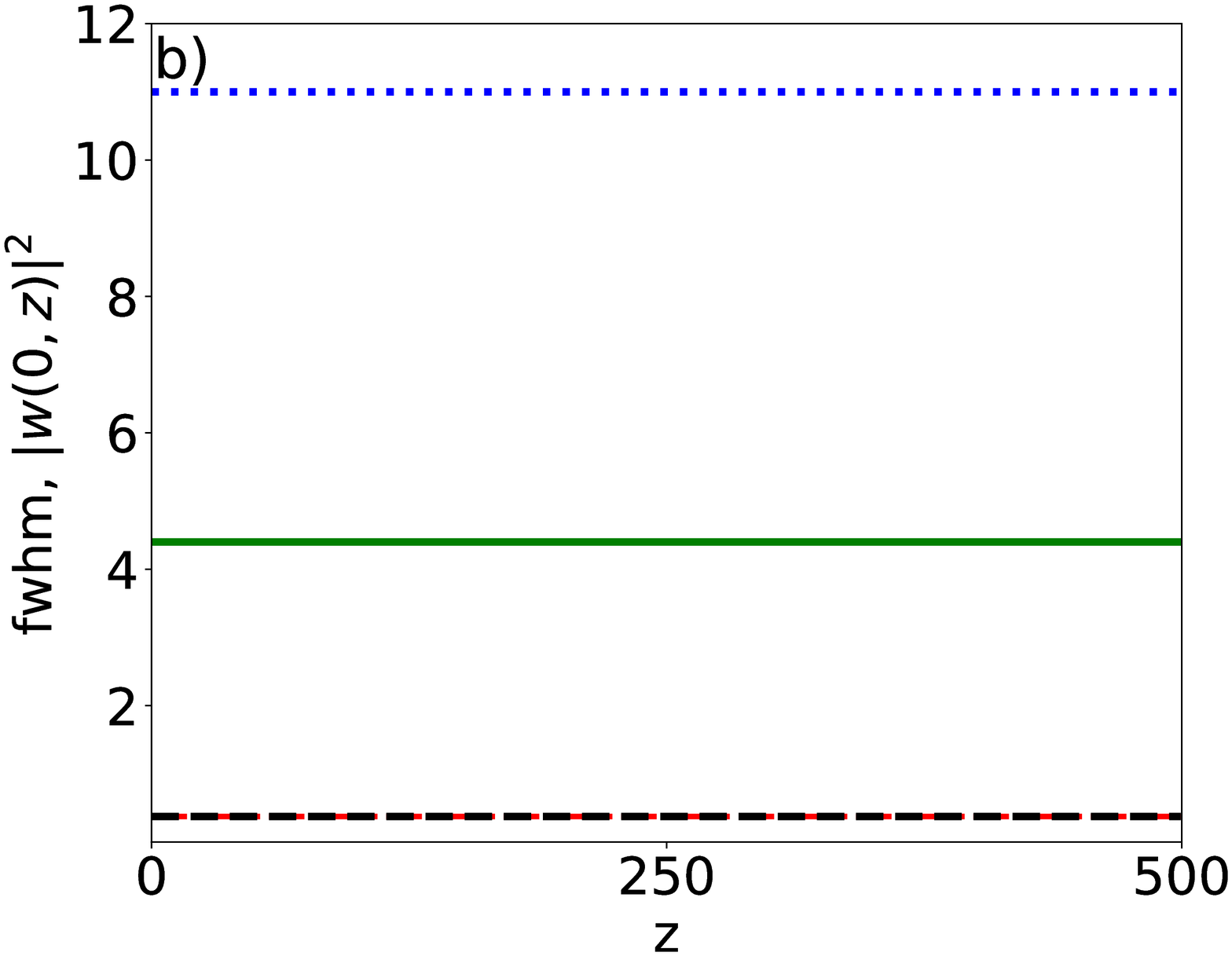}
\caption{(Color online) Evolution of width and the squared central amplitude for stationary solutions with and without modulation of the mismatch parameter $q$ in the case of large mismatch parameter, $q_0=22$. Frame a) demonstrate the unstable dynamics of the $V$ - type soliton in the case of $\chi_0=1$, dash-dotted (red) and dotted (blue) curves correspond to the soliton's full width at half maximum (\textit{fwhm}) and the squared central amplitude ($|w(0,z)|^2$), respectively, of the second-harmonic; and for $\chi_0=2.5$, dashed (black) and solid (green) curves correspond to soliton's full width at half maximum (\textit{fwhm}) and the squared central amplitude ($|w(0,z)|^2$), respectively, when the mismatch parameter is not modulated ($q_1=0$). Frame b) corresponds to stabilization of the solitons for the same stationary solutions when the mismatch parameter $q(z)$ of Eq.~(\ref{modulated_parameters}) is modulated with the frequency $\omega=9$ ($q_1=46$: red and blue curves, and $q_1=36$: black and green).}
\label{fig:5}
\end{center}
\end{figure}

For the case of a large mismatch parameter we carried out numerical calculations of the original equations~(\ref{model}), for the parameter $q_0=22$ and the maximum propagation distance, $z$, that the simulations are performed for is $z_{max}=500$. In this case, according to our numerical analysis, $V$ - type solitons with power $P_v=4\sqrt{22}$ are unstable in the region $0 < \chi_0 \lessapprox 2.8$. However, according to Eqs.~(\ref{averaged_equations_2}), this unstable region can be shifted to the stable region by modulating the mismatch parameter. Fig.~\ref{fig:5} shows the stability of solitons for the cases of $\chi_0=1$, and $\chi_0=2.5$, when the mismatch parameter are modulated with amplitudes $q_1=46$, and $q_1=36$, respectively.

\section{Modulation of the Kerr nonlinearity}
In this section, we deal with the system of a constant mismatch parameter ($q = q_0$) and a longitudinally modulated cubic nonlinearity, $\chi \rightarrow \chi(z)$. 
To derive the corresponding average over the rapid modulation of the original system~(\ref{model}), we will use the following transformations to new fields for the FH and SH
\begin{equation}\label{transformations_Kerr}
u=\bar{u}e^{i\varGamma(z)(\frac{1}{2\sigma}\lvert \bar{u}\rvert^2 + \rho\lvert \bar{v}\rvert^2)}, \: w=\bar{v}e^{i\frac{\varGamma(z)}{\sigma}(2\sigma\lvert \bar{v}\rvert^2 + \rho\lvert \bar{u}\rvert^2)},
\end{equation}
where $\varGamma(z)$ is the anti-derivative of $\chi_1(z)$, i.e., $\varGamma_z(z)=\chi_1(z)$. 
For convenience, we will drop the bar sign in the subsequent calculations. 
Inserting the transformations (\ref{transformations_Kerr}) into the original equations (\ref{model}) allows us to exclude the strongly and rapidly varying terms from the original system and derive an averaged system~\cite{Zhar}. Then we can replace the original equations (\ref{model}) with the equivalent equations:

\begin{equation}\label{sys2_with_transformation_1}
\begin{split}
iu_z & -\varGamma(z)\left(\frac{1}{2\sigma}\lvert u\rvert^2 + \rho\lvert v\rvert^2\right) _z u + \frac{\partial^2 u}{\partial x^2} + i\varGamma(z)\left(\frac{1}{2\sigma}\lvert u\rvert^2 + \rho\lvert v\rvert^2\right)_{xx}u  \\ & + 2i\varGamma(z) \left(\frac{1}{2\sigma}\lvert u\rvert^2 + \rho\lvert v\rvert^2\right)_x u_x - \varGamma^2(z)\left[\left(\frac{1}{2\sigma}\lvert u\rvert^2 + \rho\lvert v\rvert^2\right)_x \right]^2u - u \\
& + u^*ve^{i\frac{\varGamma(z)}{\sigma}\left[2\sigma(1 - \rho)\lvert v\rvert^2 + (\rho-1)\lvert u\rvert^2\right]}  + \chi_0 \left(\frac{1}{2\sigma}\lvert u\rvert^2 + \rho\lvert v\rvert^2\right)u=0,
\end{split}
\end{equation}

\begin{equation}\label{sys2_with_transformation_2}
\begin{split}
iv_z & -\varGamma(z)\left(2\sigma \lvert v\rvert^2 + \rho\lvert u\rvert^2 \right) _z v + \frac{\partial^2 v}{\partial x^2} + i\frac{\varGamma(z)}{\sigma}\left(2\sigma \lvert v\rvert^2 + \rho\lvert u\rvert^2 \right)_{xx}v \\ & + i\frac{2\varGamma(z)}{\sigma}\left(2\sigma \lvert v\rvert^2 + \rho\lvert u\rvert^2 \right)_x v_x - \frac{\varGamma^2(z)}{\sigma^2}\left[\left(2\sigma \lvert v\rvert^2 + \rho\lvert u\rvert^2 \right)_x \right]^2v - qv \\ & + \frac{1}{2} u^2e^{-i\frac{\varGamma(z)}{\sigma}\left[2\sigma(1 - \rho)\lvert v\rvert^2 + (\rho-1)\lvert u\rvert^2\right]} + \chi_0\left(2\sigma \lvert v\rvert^2 + \rho\lvert u\rvert^2 \right)v=0.
\end{split}
\end{equation}
It follows from Eqs.~(\ref{sys2_with_transformation_1}) and (\ref{sys2_with_transformation_2}) that
\begin{equation}\label{z_derivatives_1}
\begin{split}
& \left(\frac{1}{2\sigma}\lvert u\rvert^2 + \rho\lvert v\rvert^2\right) _z = \frac{-i}{2\sigma}uu_{xx}^*+\frac{i}{2\sigma}u^*u_{xx} - \frac{\varGamma(z)}{\sigma}\left[\left(\frac{1}{2\sigma}\lvert u\rvert^2 + \rho\lvert v\rvert^2\right)_x\lvert u\rvert^2\right]_x \\
&- \frac{i(\rho-1)}{2\sigma} u^{*2}ve^{i\frac{\varGamma(z)}{\sigma}\left[2\sigma(1 - \rho)\lvert v\rvert^2 + (\rho-1)\lvert u\rvert^2\right]} +\frac{i(\rho-1)}{2\sigma} u^2v^*e^{-i\frac{\varGamma(z)}{\sigma}\left[2\sigma(1 - \rho)\lvert v\rvert^2 + (\rho-1)\lvert u\rvert^2\right]} \\ &- \frac{i\rho}{\sigma}vv_{xx}^*+\frac{i\rho}{\sigma}v^*v_{xx} - \frac{2\rho}{\sigma} \frac{\varGamma(z)}{\sigma}\left[\left(2\sigma \lvert v\rvert^2 + \rho\lvert u\rvert^2 \right)_x\lvert v\rvert^2\right]_x,
\end{split}
\end{equation}
and
\begin{equation}\label{z_derivatives_2}
\begin{split}
&\left(2\sigma \lvert v\rvert^2 + \rho\lvert u\rvert^2 \right) _z = -i\rho uu_{xx}^*+i\rho u^*u_{xx} - 2\rho \varGamma(z)\left[\left(\frac{1}{2\sigma}\lvert u\rvert^2 + \rho\lvert v\rvert^2\right)_x\lvert u\rvert^2\right]_x \\
& + i(\rho-1) u^{*2}ve^{i\frac{\varGamma(z)}{\sigma}\left[2\sigma(1 - \rho)\lvert v\rvert^2 + (\rho-1)\lvert u\rvert^2\right]} -i(\rho-1) u^2v^*e^{-i\frac{\varGamma(z)}{\sigma}\left[2\sigma(1 - \rho)\lvert v\rvert^2 + (\rho-1)\lvert u\rvert^2\right]} \\
&- i2vv_{xx}^*+i2v^*v_{xx} - \frac{4\varGamma(z)}{\sigma}\left[\left(2\sigma \lvert v\rvert^2 + \rho\lvert u\rvert^2 \right)_x\lvert v\rvert^2\right]_x.
\end{split}
\end{equation}

We insert Eq.~(\ref{z_derivatives_1}) and Eq.~(\ref{z_derivatives_2}) into Eqs.~(\ref{sys2_with_transformation_1}) and (\ref{sys2_with_transformation_2}), then we average over the period $\Lambda=2\pi/\omega$ for the rapid oscillations, using the relations
\begin{eqnarray}\label{bessels}
\langle \varGamma(z) \rangle = 0, \,\,\,\, \langle \varGamma^2(z) \rangle = \beta^2, \nonumber \\
\langle e^{\pm i\frac{\varGamma(z)}{\sigma}\Theta}\rangle=\frac{1}{\Lambda}\int_0^{\Lambda}e^{\pm i\frac{\varGamma(z)}{\sigma}\Theta}dz=J_0 \left( \frac{\chi_1}{\sigma\omega}\Theta \right), \nonumber \\
\langle \varGamma(z) e^{\pm i\frac{\varGamma(z)}{\sigma}\Theta}\rangle=\frac{1}{\Lambda}\int_0^{\Lambda} \varGamma(z) e^{\pm i\frac{\varGamma(z)}{\sigma}\Theta}dz \nonumber \\
=\pm i\frac{\chi_1}{\omega} J_1 \left( \frac{\chi_1}{\sigma\omega}\Theta \right), \nonumber
\end{eqnarray}
where $\Theta = \left[(1 - \rho)(2\sigma\lvert v\rvert^2 - \lvert u\rvert^2\right)]$ and $J_i(\cdot), i=0,1$ are the zero- and first-order Bessel functions.
In the averaging process fast oscillating terms (terms  with $\varGamma(z)$) vanish. We obtain the following system of averaged coupled equations for the FH:
\begin{equation}\label{averaged_equations_u_3}
\begin{split}
& iu_z  + \beta^2 \left[ \frac{1}{\sigma}\left[\left(\frac{1}{2\sigma}\lvert u\rvert^2 + \rho\lvert v\rvert^2\right)_x\lvert u\rvert^2\right]_x + \frac{2\rho}{\sigma^2}\left[\left(2\sigma \lvert v\rvert^2 + \rho\lvert u\rvert^2 \right)_x\lvert v\rvert^2\right]_x\right]u\\
& - \beta^2 \left[\left(\frac{1}{2\sigma}\lvert u\rvert^2 + \rho\lvert v\rvert^2\right)_x\right]^2 u
 +\frac{\partial^2 u}{\partial x^2} - \frac{\chi_1}{\omega}\frac{\rho-1}{2\sigma} J_1 \left( \frac{\chi_1}{\sigma\omega}\Theta \right)u\left(u^2v^* + u^{*2}v \right)\\
 & -u + u^*vJ_0 \left( \frac{\chi_1}{\sigma\omega}\Theta \right) + \chi_0 \left(\frac{1}{2\sigma}\lvert u\rvert^2 + \rho\lvert v\rvert^2 \right)u=0, \\
\end{split}
\end{equation}
and SH:
\begin{equation}\label{averaged_equations_v_3}
\begin{split}
& i\sigma v_z  + \beta^2 \left[ 2\rho\left[\left(\frac{1}{2\sigma}\lvert u\rvert^2 + \rho\lvert v\rvert^2\right)_x\lvert u\rvert^2\right]_x + \frac{4}{\sigma}\left[\left(2\sigma \lvert v\rvert^2 + \rho\lvert u\rvert^2 \right)_x\lvert v\rvert^2\right]_x\right]v \\
& - \frac{\beta^2}{\sigma^2}\left[\left(2\sigma \lvert v\rvert^2 + \rho\lvert u\rvert^2 \right)_x\right]^2 v +\frac{\partial^2 v}{\partial x^2} + \frac{\chi_1}{\omega}(\rho-1) J_1 \left( \frac{\chi_1}{\sigma\omega}\Theta \right)v\left(u^2v^* +u^{*2}v\right) \\
& -qv +\frac{1}{2}u^2J_0 \left( \frac{\chi_1}{\sigma\omega}\Theta \right) + \chi_0(2\sigma\lvert v\rvert^2 + \rho\lvert u\rvert^2)v=0.
\end{split}
\end{equation}

The Hamiltonian of the above averaged system, i.e.
$$iu_z=\frac{\delta \langle H \rangle}{\delta u^*}, \,\,\,\ i\sigma v_z=\frac{\delta \langle H \rangle}{\delta v^*},$$
is then
\begin{equation} \label{Averaged_Hamiltonian}
\begin{split}
\langle H\rangle = & \int_{-\infty}^{+\infty} \bigg[\lvert u_x \rvert^2+\lvert v_x \rvert^2+ \beta^2 \left[\left(\frac{1}{2\sigma}\lvert u\rvert^2 + \rho\lvert v\rvert^2 \right)_x \right]^2 \lvert u \rvert^2  \\
&+ \frac{\beta^2}{\sigma^2}\left[\left(2\sigma \lvert v\rvert^2 + \rho\lvert u\rvert^2\right)_x\right]^2 \lvert v \rvert^2 - \frac{\chi_0}{4\sigma}\lvert u\rvert^4 -\chi_0\sigma\lvert v\rvert^4 + q\lvert v \rvert^2 + \lvert u \rvert^2 \\
&- \chi_0 \rho \lvert v \rvert^2 \lvert u \rvert^2 - \frac{1}{2} J_0 \left( \frac{\chi_1}{\sigma\omega}\Theta \right) (u^2v^*+u^{*2}v)\bigg] dx.
\end{split}
\end{equation}

The effective $\chi^{(2)}_{\textnormal{eff}}$ non-linearity parameter now become $\chi^{(2)}_{\textnormal{eff}}=\chi^{(2)} J_0 \left( \frac{\chi_1}{\sigma\omega}\Theta \right)$, i.e. it depends also on the power imbalance of FH and SH. In the case of AMBEC, it corresponds to the dependence of $\chi^{(2)}$ on the population imbalance between the atoms and the molecular BEC~\cite{Magnus}.

\begin{figure}[H]
\begin{center}
\includegraphics[scale=0.27]{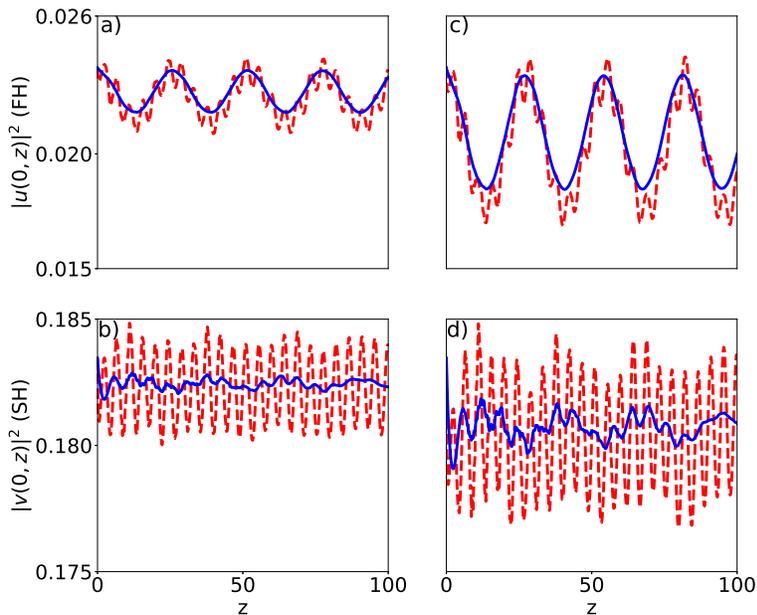}
\caption{(Color online) Evolution of stationary solutions in the cases of two different modulation parameters for the Kerr nonlinearity. In the left-hand side, frames a) and b), we demonstrate the evolution of the squared central amplitude for the fundamental- and second-harmonics ($|u(0,z)|^2$ and $|v(0,z)|^2$, respectively) in the case of $\chi_1 = 6$. On the right-hand side, frames c) and d), corresponds to the case of $\chi_1=10$. Dashed (red) and solid (blue) curves correspond to the original model, Eqs.~(\ref{model}), and the averaged equations, Eqs.~(\ref{averaged_equations_u_3}) and (\ref{averaged_equations_v_3}), respectively.} 
\label{fig:6}
\end{center}
\end{figure}

To check the validity of this averaged model, we perform numerical simulation of the original system~(\ref{model}), and the averaged equations, Eqs.~(\ref{averaged_equations_u_3}) and (\ref{averaged_equations_v_3}), with the same initial condition obtained by solving numerically for the stationary states of Eqs.~(\ref{model}) with the parameters $q=2$, $\chi_0=5$. 
This comparison is reported in Fig.~\ref{fig:6} for evolution of the stationary solutions. With the modulation parameters $\omega=30$, $\chi_1 = 6$ and $\chi_1=10$, we have confirmed agreement between the results from the original and the averaged models.

\section{Conclusion}\label{conc}
In conclusion, we suggest the method of dynamical stabilization of solitons in media with competing quadratic and cubic nonlinearities, based on the rapid modulations of the longitudinal  variable of the mismatch parameter. The criteria for the dynamical stabilization of solitons for mismatch management are obtained from analysis of averaged equations. In addition, in the case of the management of the cubic nonlinearity, we have shown that stable optical solitons can exist in medium supported by competing quadratic and varying Kerr non-linearities. It is shown in the analysis of an averaged system, that the Kerr non-linearity management is described by the same non-modulated Hamiltonian but with an effective $\chi^{(2)}_{\textnormal{eff}}$ parameter that depends on the intensity imbalance between the FH and SH fields. Finally, we have confirmed that the predictions of the average model corroborate with the full numerical simulations.

\section*{Acknowledgments}
This work has been supported by the state budget of the Republic of Uzbekistan, and for M. \"{O}. through ORU-RR-2021/2022. We also thank the authors of the XMDS
software~\cite{xmds}, which was used here for the dynamical simulations.


\end{document}